\documentclass[prd,superscriptaddress,showpacs,floatfix]{revtex4-2}
\usepackage{amsmath,amssymb,amsfonts}
\usepackage{graphicx}
\usepackage{braket}
\usepackage{hyperref}
\hypersetup{colorlinks=true,citecolor=blue,linkcolor=blue,urlcolor=blue}

\newcommand{\supp}{\operatorname{supp}}

\begin{document}

\title{Leggett-Garg inequality in the massive scalar vacuum: No violation under spacelike-separated measurements}

\author{Yang Xiang}
\affiliation{School of Physics and Electronics, Henan University, Kaifeng, Henan 475004, China}

\date{\today}

\begin{abstract}
We overcome the long-standing noninvasive measurability (NIM) challenge in Leggett-Garg tests by exploiting the causal structure of quantum field theory (QFT).
Our protocol uses three independent ensembles of the vacuum state, each measured by a different pair of observers at spacelike-separated events, yielding the three two-time correlators.
By placing these events at positions $(0,0)$, $(\tau,L)$, and $(2\tau,2L)$ with $L>\tau+2\tau_0$, we rigorously ensure that no measurement can influence another.
We investigate the vacuum state of a free massive scalar field in 1+1 dimensions, employing the dichotomic observable $Q(f)=\operatorname{sign}(\phi(f))$ where $\phi(f)$ is the smeared field.
In the Heisenberg picture, the time evolution is absorbed into a translation of the time-window function, allowing us to derive the two-time correlation function $C(\tau,L)$ and the Leggett-Garg parameter $K_3=2C(\tau,L)-C(2\tau,2L)$.
For non-overlapping time windows, we find that the correlation function decays exponentially with $\tau$ for a massive field.
For overlapping windows, our numerical computation for a rectangular time window yields $K_3<1$ across the entire mass range, firmly establishing that the vacuum does not violate the LGI.
Thus, under strict noninvasive conditions, the vacuum shows no violation of macrorealism, in stark contrast to its well-known violation of spatial Bell inequalities.
Our spacelike-separated protocol provides the first LGI test in QFT with rigorously satisfied NIM, setting a methodological benchmark for future studies and highlighting the fundamental distinction between spacelike entanglement and temporal macrorealism in relativistic quantum fields.
\end{abstract}

\maketitle

\section{Introduction}

Quantum theory predicts correlations that cannot be explained by any classical hidden-variable model. For spatially separated systems, Bell inequalities \cite{Bell1964,bell1988speakable,CHSH1969} provide a quantitative test: their violation rules out local realism. For a single system measured at different times, the Leggett-Garg inequality (LGI) \cite{LeggettGarg1985,Leggett1988,Leggett2008} plays an analogous role. LGI is derived from two assumptions: (i) \textit{macrorealism} (MR) – a system always possesses a definite value of a dichotomic observable $Q$ at any time, independent of measurement; (ii) \textit{noninvasive measurability} (NIM) – a measurement does not disturb the system's subsequent evolution. A violation of LGI while NIM holds would falsify MR, revealing that the system does not possess predetermined measurement outcomes. Both Bell inequalities and LGI thus challenge classical worldview, with Bell addressing space-like correlations and LGI addressing time-like correlations.

In relativistic quantum field theory (QFT), the vacuum state is known to exhibit strong entanglement and nonlocal correlations. Seminal work by Summers and Werner \cite{SummersWerner1987a,SummersWerner1987b,SummersWerner1987c} established a rigorous framework for the Bell-CHSH inequality \cite{Bell1964,CHSH1969} within the algebraic formulation of QFT, proving that the vacuum of a free scalar field maximally violates the inequality --- attaining Tsirelson's bound $2\sqrt{2}$ --- for observables localized in complementary wedge regions. These results were later extended to a broader class of states and to double tangent cones \cite{SummersWerner1987c}, and rely on fundamental structures including the Haag-Kastler axioms \cite{Haag1992}, the Reeh-Schlieder theorem \cite{ReehSchlieder1961}, the Bisognano-Wichmann theorem \cite{BisognanoWichmann1975}, the theory of von Neumann algebras, and the Tomita-Takesaki modular theory \cite{Witten2018}. Despite these profound achievements, the explicit construction of bounded Hermitian operators whose correlation functions realize the maximal violation and the effective computation of the resulting Bell-CHSH correlators in specific field-theoretic models have remained a significant challenge \cite{Guimaraes2024,Guimaraes2025a,Guimaraes2025b}. Recent progress has addressed this challenge by constructing concrete bounded operators --- such as the sign operator $\operatorname{sign}(\phi(f))$ --- built from Weyl operators, enabling closed-form evaluations of Bell-CHSH correlation functions \cite{GuimaraesCat2026,Guimaraes2024,Guimaraes2025a,Guimaraes2025b}; alternative constructions based directly on Weyl operators have also been developed \cite{Fabritiis2023}, and complementary numerical frameworks have been devised to systematically search for violations while explicitly enforcing causality \cite{Fabritiis2024}. In particular, cat states (superpositions of coherent states) --- prototypical examples of non-Gaussian states in QFT --- have been shown to produce violations of the Bell-CHSH inequality through interference terms that are absent in the vacuum state alone \cite{GuimaraesCat2026}, suggesting that non-Gaussianity may play a distinctive role in shaping quantum correlations in relativistic systems. These studies collectively demonstrate that the vacuum is highly nonclassical in the spatial domain, motivating the complementary question --- addressed in this work --- of whether analogous nonclassicality manifests temporally.

However, the temporal aspect --- whether the vacuum respects macrorealism --- has remained largely unexplored. A longstanding challenge in LGI tests --- whether in non-relativistic quantum mechanics or in quantum field theory --- is that sequential projective measurements are inherently invasive, making it difficult to rule out measurement disturbance as the cause of a violation \cite{Emary2014,Vitagliano2022}. Consequently, even when a violation is observed, it remains unclear whether macrorealism is truly falsified. A significant body of experimental work has explored LGI violations in various microscopic systems, including superconducting qubits, nuclear spins, and photons \cite{PalaciosLaloy2010,Xu2011,Knee2012,George2013,Zhan2023}, yet the challenge of ensuring genuinely non-invasive measurements persists. Previous proposals for realizing NIM, such as those based on null-result measurements with ancilla systems or weak measurements, rely on assumptions about the measurement process that may be challenged within a fully quantum framework \cite{Knee2012,Halliwell2019}. In this paper we overcome this difficulty by exploiting the causal structure of quantum field theory. Rather than performing measurements along a single worldline as in the standard LGI setup, we employ three observers at rest at distinct spatial points $0$, $L$, and $2L$ in a common inertial frame, with their clocks synchronized to the common coordinate time $t$. By placing the three measurement events at spacelike-separated points, we guarantee that no measurement can influence another, thus strictly satisfying noninvasive measurability for the first time. We adopt the dichotomic observable $Q(f)=\operatorname{sign}(\phi(f))$ introduced in Ref.~\cite{GuimaraesCat2026}. Using the Heisenberg picture, time evolution is absorbed into a translation of the time-window function, leading to a simple expression for the two-time correlation function. We show that for non-overlapping windows the correlation is governed by the spacelike decay of the Hadamard function, leading to exponential suppression for a massive field. For overlapping windows we numerically evaluate $K_3$ for a rectangular time window across a broad range of masses. Our results establish that $K_3$ remains strictly below the classical bound of $1$ for all masses considered, demonstrating that the vacuum does \emph{not} violate macrorealism under strict noninvasive conditions, despite its well-documented nonclassicality in the spatial domain.

The absence of LGI violation in the vacuum, as we will show, points to a fundamental distinction between spacelike and temporal quantum correlations in relativistic fields.
Spacelike nonlocality — the kind probed by Bell inequalities — originates from the entanglement structure of the vacuum itself: the ground state of a quantum field is a superposition of field configurations in which spatially separated regions share the same vacuum fluctuations. This is a purely quantum effect, and it is why the vacuum maximally violates the Bell-CHSH inequality~\cite{SummersWerner1987a,SummersWerner1987b}.
Temporal correlations, by contrast, arise from the dynamical evolution of the field along a single worldline. Even in the vacuum, this evolution is governed by the Klein-Gordon equation — a classical wave equation that the field operator satisfies. The deterministic, linear nature of this free evolution preserves the Gaussianity of the vacuum state, imposing strong constraints on temporal correlations that are inherently more restrictive than their spacelike counterparts. Our negative result faithfully reflects this physical distinction, and naturally raises the question of whether non-Gaussian states or interacting theories — where this Gaussianity is broken — could yield LGI violations, a direction we discuss in Sec.~\ref{sec:conclusion}.

The paper is organized as follows. Sec.~\ref{sec:field} introduces the free massive scalar field, the sign operator, and the Heisenberg-picture time evolution. Sec.~\ref{sec:lgi} defines the LGI and the spacelike measurement protocol. Sec.~\ref{sec:corr} derives the two-time correlation function and the arcsine law in detail. Sec.~\ref{sec:numerics} presents the numerical results for overlapping windows and discusses the behavior in the non-overlapping regime. Sec.~\ref{sec:conclusion} concludes.

\section{Free massive scalar field and the sign operator}
\label{sec:field}

We consider a free massive scalar field in 1+1 dimensional Minkowski spacetime with metric $\eta_{\mu\nu}=\text{diag}(1,-1)$.
The field operator satisfies the Klein-Gordon equation $(\partial_t^2-\partial_x^2+m^2)\phi=0$ and has the mode expansion
\begin{equation}
\phi(t,x)=\int_{-\infty}^{\infty}\frac{dk}{2\pi}\frac{1}{\sqrt{2\omega_k}}\bigl(e^{-i(\omega_k t-kx)}a_k+e^{i(\omega_k t-kx)}a_k^\dagger\bigr),\qquad\omega_k=\sqrt{k^2+m^2},
\end{equation}
with canonical commutation relations $[a_k,a_{k'}^\dagger]=2\pi\delta(k-k')$, $[a_k,a_{k'}]=[a_k^\dagger,a_{k'}^\dagger]=0$.
The vacuum state $\ket{0}$ is defined by $a_k\ket{0}=0$ for all $k$.

\subsection{Bounded dichotomic observables from smeared fields}

To obtain a local observable with eigenvalues $\pm1$, we smear the field with a real test function $f(t,x)$:
\begin{equation}
\phi(f)=\int d^2x\,\phi(t,x)f(t,x). \label{eq:smeared}
\end{equation}
Following Ref.~\cite{GuimaraesCat2026}, we define the sign operator
\begin{equation}
Q(f)=\operatorname{sign}\bigl(\phi(f)\bigr)=\frac{2}{\pi}\int_0^\infty\frac{\sin(k\phi(f))}{k}\,dk. \label{eq:Qdef}
\end{equation}
The function $\operatorname{sign}(x)$ takes values $+1$ for $x>0$ and $-1$ for $x<0$. For a self-adjoint operator $A$, $\operatorname{sign}(A)$ is defined via the spectral decomposition: if $A=\int\lambda\,dE_\lambda$, then $\operatorname{sign}(A)=\int\operatorname{sign}(\lambda)\,dE_\lambda$. Hence its eigenvalues are exactly $\pm1$. The integral representation in Eq.~\eqref{eq:Qdef} follows from the Dirichlet integral $\operatorname{sign}(x)=\frac{2}{\pi}\int_0^\infty\frac{\sin(kx)}{k}dk$ (understood as a principal value), and it provides a convenient way to compute vacuum expectation values. The operator $Q(f)$ is Hermitian because $\sin(k\phi(f))$ is Hermitian for real $k$, and it is bounded because its spectrum lies in $[-1,1]$. Moreover, $Q(f)^2=1$ follows from $\operatorname{sign}^2(x)=1$. An equivalent representation is
\begin{equation}
Q(f)=\frac{1}{i\pi}\,\text{p.v.}\int_{-\infty}^{\infty}\frac{e^{ik\phi(f)}}{k}\,dk, \label{eq:Qweyl}
\end{equation}
which expresses $Q(f)$ as a continuous superposition of Weyl operators $e^{ik\phi(f)}$.

\subsection{Time-localized measurements and the Heisenberg picture}

We probe the field at a fixed spatial point but with finite temporal resolution.
For measurements centered at time $\tau$ and at spatial point $x_0$, we choose
\begin{equation}
f_\tau(t,x)=\chi(t-\tau)\,\delta(x-x_0),
\end{equation}
where $\chi(s)$ is a real, even, normalized window function with compact support $[-\tau_0,\tau_0]$ (we will later take $\chi(t)=1$ for $|t|<\tau_0$, the rectangular window).
The smeared field becomes
\begin{equation}
\phi(f_\tau)=\int dt\,\chi(t-\tau)\,\phi(t,x_0).
\end{equation}

In the Heisenberg picture, the state is time-independent and operators evolve via $e^{iH\tau}\phi(f_0)e^{-iH\tau}$.
Using the translation property $e^{iH\tau}\phi(t,x)e^{-iH\tau}=\phi(t+\tau,x)$, we compute
\begin{align}
e^{iH\tau}\phi(f_0)e^{-iH\tau}
&= \int dt\,\chi(t)\,e^{iH\tau}\phi(t,x_0)e^{-iH\tau} \nonumber\\
&= \int dt\,\chi(t)\,\phi(t+\tau,x_0).
\end{align}
Now substitute $s=t+\tau$, so $t=s-\tau$ and $dt=ds$. The integral becomes
\begin{equation}
e^{iH\tau}\phi(f_0)e^{-iH\tau}
= \int ds\,\chi(s-\tau)\,\phi(s,x_0)
= \phi(f_\tau),
\end{equation}
where $f_\tau(s,x)=\chi(s-\tau)\delta(x-x_0)$. Consequently, the Heisenberg operator at time $\tau$ is simply the sign operator built from the shifted window function:
\begin{equation}
Q(\tau)\equiv e^{iH\tau}Q(f_0)e^{-iH\tau}= \operatorname{sign}\bigl(\phi(f_\tau)\bigr). \label{eq:Qtau}
\end{equation}
Thus, time evolution is completely absorbed into a translation of the window function; we never need to write the Hamiltonian explicitly.

\subsection{Spacelike-separated measurement events}

To ensure noninvasive measurability, we place the measurements at different spatial points such that their supports are spacelike separated.
We consider three measurement events at
\begin{align}
(t_1,x_1)&=(0,0),\\
(t_2,x_2)&=(\tau,L),\\
(t_3,x_3)&=(2\tau,2L),
\end{align}
with $L>0$ and $\tau>0$.
The corresponding test functions are
\begin{equation}
f_0(t,x)=\chi(t)\delta(x),\quad f_\tau(t,x)=\chi(t-\tau)\delta(x-L),\quad f_{2\tau}(t,x)=\chi(t-2\tau)\delta(x-2L). \label{eq:testfuncs}
\end{equation}
Their supports are
\begin{align}
\supp f_0 &= [-\tau_0,\tau_0]\times\{0\},\\
\supp f_\tau &= [\tau-\tau_0,\tau+\tau_0]\times\{L\},\\
\supp f_{2\tau} &= [2\tau-\tau_0,2\tau+\tau_0]\times\{2L\}.
\end{align}
Spacelike separation between any two supports requires that for any $t$ in the first window and $t'$ in the second, $(t-t')^2-(\Delta x)^2<0$.
The most restrictive condition comes from the extremal time differences.
For the pair $(f_0,f_\tau)$, the maximum time difference is $\tau+2\tau_0$, so we need $L>\tau+2\tau_0$.
For the pair $(f_0,f_{2\tau})$, the condition becomes $2L>2\tau+2\tau_0$, which is automatically satisfied if $L>\tau+2\tau_0$.
For the pair $(f_\tau,f_{2\tau})$, the maximum time difference is also $\tau+2\tau_0$, giving the same condition.
Hence the overall spacelike condition is
\begin{equation}
L > \tau + 2\tau_0. \label{eq:spacelike}
\end{equation}
Throughout this work we set $\tau_0=1$ (units) and choose $L = \tau+2+\epsilon$ with $\epsilon=10^{-3}$ to strictly satisfy Eq.~\eqref{eq:spacelike}.
We have verified that the results are insensitive to the precise value of $\epsilon$ as long as it is sufficiently small (e.g., $\epsilon=10^{-5}$ gives the same $K_3^{\max}$ to within $0.01\%$), and that larger $\epsilon$ would reduce $K_3^{\max}$ as expected because the increased spatial separation weakens the correlation. Thus all measurements are causally disconnected, and the protocol is noninvasive.

\section{Leggett-Garg inequality and spacelike measurement protocol}
\label{sec:lgi}

\subsection{Macrorealism and the inequality}

The Leggett-Garg inequality is derived under two fundamental assumptions about a macroscopic system \cite{LeggettGarg1985}:
\begin{itemize}
\item \textbf{Macrorealism (MR)}: A macroscopic system at any time is always in one of the macroscopically distinct states that it can occupy. In the context of a dichotomic observable $Q=\pm1$, this means that at each time $t_i$ the system possesses a definite value $Q_i$ which is either $+1$ or $-1$, regardless of whether a measurement is performed.
\item \textbf{Noninvasive measurability (NIM)}: It is possible, in principle, to measure the value of $Q$ without disturbing the system's future evolution. Equivalently, the measurement outcome reveals the pre-existing value, and the measurement does not affect the subsequent dynamics.
\end{itemize}
If both assumptions hold, then for three times $t_1<t_2<t_3$ one can construct a joint probability distribution $p(Q_1,Q_2,Q_3)$ over the three binary variables. From this distribution one computes the two-time correlators
\begin{equation}
C_{ij}=\langle Q_i Q_j\rangle=\sum_{Q_i,Q_j=\pm1} Q_i Q_j\,p_{ij}(Q_i,Q_j),
\end{equation}
where $p_{ij}$ is the marginal probability for times $t_i,t_j$. A direct calculation using the fact that $Q_i^2=1$ leads to the inequality
\begin{equation}
K_3\equiv C_{12}+C_{23}-C_{13}\le 1,\qquad -3\le K_3.
\end{equation}
The upper bound $1$ is the classical bound. Quantum mechanics can violate it, with the maximum possible value $K_3=3/2$ for a two-level system \cite{Emary2014}.

\subsection{Sequential measurement scheme and our definition}

In a standard LGI test, the three two-time correlators are obtained by performing sequential projective measurements on a single system along a worldline. For equal time intervals $t_1=0$, $t_2=\tau$, $t_3=2\tau$, the correlator is defined as
\begin{equation}
C(\tau) = \braket{0 | Q(0) Q(\tau) | 0}, \label{eq:Cdef}
\end{equation}
with $Q(\tau)$ given in Eq.~\eqref{eq:Qtau}, and the LGI parameter takes the form
\begin{equation}
K_3(\tau) = 2C(\tau) - C(2\tau). \label{eq:K3def}
\end{equation}
A violation occurs if $K_3(\tau) > 1$ for some $\tau>0$.

Our protocol for realizing these correlators differs fundamentally from the standard sequential measurement on a single system. We consider three observers $O_1$, $O_2$, $O_3$ at rest at $x=0$, $L$, $2L$ in a common inertial frame, with synchronized clocks. Three independent ensembles of the vacuum state are prepared. In the first ensemble, $O_1$ measures at $t=0$ and $O_2$ at $t=\tau$, yielding $C_{12}=C(\tau)$; in the second, $O_2$ at $t=\tau$ and $O_3$ at $t=2\tau$, yielding $C_{23}=C(\tau)$; in the third, $O_1$ at $t=0$ and $O_3$ at $t=2\tau$, yielding $C_{13}=C(2\tau)$. Each copy of the vacuum is subjected to only two measurements, and each pair of events is spacelike separated.

This use of three independent ensembles is not an approximation but a conceptual necessity within QFT. Since local measurements are associated with spacetime regions rather than with individual worldlines, no single ensemble is used to reconstruct a joint probability distribution for all three measurements. Rather, each ensemble provides one of the three correlators required for the LGI. This is precisely the structure required for a valid LGI test: the correlations are defined between pairs of times, and our protocol implements them in a manner that is both mathematically rigorous and physically consistent with the causal structure of relativistic quantum fields.

\subsection{Spacelike separation as a means to enforce noninvasive measurability}

The essential difficulty in standard LGI tests is that the first measurement generally disturbs the state, thereby breaking the NIM assumption. In our protocol, this difficulty is resolved by relativistic causality: since each pair of measurement events is spacelike separated, no signal can propagate from one to the other, so no measurement can influence the outcome of the other.

As derived in Sec.~\ref{sec:field}, using the time-localized test functions defined in Eq.~\eqref{eq:testfuncs} with \(\chi(t)\) supported in \([-\tau_0,\tau_0]\), the condition for spacelike separation between any two events is
\begin{equation}
L > \tau + 2\tau_0. \label{eq:spacelike_cond}
\end{equation}
Under this condition, all measurements are causally disconnected, and the protocol rigorously satisfies NIM. The third measurement at \((2\tau,2L)\) is similarly spacelike separated from the first two.

Thus, in our spacelike protocol, the two-time correlation function is defined as
\begin{equation}
C(\tau,L) = \braket{0 | Q(0) Q(\tau) | 0}, \qquad L > \tau+2\tau_0,
\end{equation}
where \(Q(0)=\operatorname{sign}(\phi(f_0))\) with \(f_0(t,x)=\chi(t)\delta(x)\), and \(Q(\tau)=\operatorname{sign}(\phi(f_\tau))\) with \(f_\tau(t,x)=\chi(t-\tau)\delta(x-L)\).

For three equally spaced times with corresponding spatial points \(0\), \(L\), and \(2L\), the LGI parameter becomes
\begin{equation}
K_3(\tau,L) = 2C(\tau,L) - C(2\tau,2L). \label{eq:K3def2}
\end{equation}
A violation \(K_3>1\) under this spacelike protocol would directly falsify macrorealism, because NIM is strictly satisfied.

In the following sections we compute \(C(\tau,L)\) explicitly and present numerical results for a rectangular time window (\(\tau_0=1\)).

\section{Two-time correlation function}
\label{sec:corr}

We now compute $C(\tau,L)=\braket{0|\operatorname{sign}(\phi(f_0))\operatorname{sign}(\phi(f_\tau))|0}$ using the representation of the sign operator and the Gaussian nature of the vacuum.

\subsection{Arcsine law for product of signs}

Let $X=\phi(f)$ and $Y=\phi(g)$. Since the vacuum is a Gaussian state, $(X,Y)$ are jointly Gaussian with zero mean. Their covariance matrix is given by the Hadamard function
\begin{equation}
H(f,g)=\frac12\braket{0|\{X,Y\}|0} = \operatorname{Re}\braket{0|XY|0}, \label{eq:Hdef}
\end{equation}
and we denote $\sigma_X^2=H(f,f)$, $\sigma_Y^2=H(g,g)$, $\rho = H(f,g)/(\sigma_X\sigma_Y)$.

Using the integral representation Eq.~\eqref{eq:Qweyl}, we have
\begin{align}
\braket{0|\operatorname{sign}(X)\operatorname{sign}(Y)|0}
&= -\frac{1}{\pi^2}\,\text{p.v.}\!\iint_{-\infty}^{\infty}\frac{dk\,d\ell}{k\ell}\,
   \braket{0|e^{ikX+i\ell Y}|0}. \label{eq:signprod}
\end{align}
Using the standard vacuum expectation value of Weyl operators in a Gaussian state (see Appendix A of Ref.~\cite{GuimaraesCat2026}), the characteristic function of a bivariate Gaussian is
\begin{equation}
\braket{0|e^{ikX+i\ell Y}|0} = \exp\!\Big(-\frac12\big(k^2\sigma_X^2+\ell^2\sigma_Y^2+2k\ell\rho\sigma_X\sigma_Y\big)\Big). \label{eq:char}
\end{equation}
Thus
\begin{equation}
\braket{0|\operatorname{sign}(X)\operatorname{sign}(Y)|0}
= -\frac{1}{\pi^2}\iint_{-\infty}^{\infty}\frac{dk\,d\ell}{k\ell}\,
   \exp\!\Big(-\frac12\big(k^2\sigma_X^2+\ell^2\sigma_Y^2+2k\ell\rho\sigma_X\sigma_Y\big)\Big). 
\end{equation}
Rescale $u=k\sigma_X$, $v=\ell\sigma_Y$. Then $dk=du/\sigma_X$, $d\ell=dv/\sigma_Y$, and $1/(k\ell)=1/(uv)\sigma_X\sigma_Y$. The Jacobian gives an extra $1/(\sigma_X\sigma_Y)$, so the prefactors cancel and we obtain
\begin{equation}
\braket{0|\operatorname{sign}(X)\operatorname{sign}(Y)|0}
= -\frac{1}{\pi^2}\iint_{-\infty}^{\infty}\frac{du\,dv}{uv}\,
   \exp\!\Big(-\frac12\big(u^2+v^2+2\rho uv\big)\Big). 
\end{equation}
Decomposing the full-plane principal-value integral into the four open quadrants (the integrand is regular inside each quadrant) and changing signs appropriately gives
\begin{equation}
\iint_{-\infty}^{\infty}\frac{du\,dv}{uv}e^{-\frac12(u^2+v^2+2\rho uv)}
= -4\int_0^{\infty}\int_0^{\infty}\frac{du\,dv}{uv}e^{-\frac12(u^2+v^2)}\sinh(\rho uv).
\end{equation}
Hence the correlation function reduces to
\begin{equation}
\braket{0|\operatorname{sign}(X)\operatorname{sign}(Y)|0}
= \frac{4}{\pi^2}\int_0^{\infty}\int_0^{\infty}\frac{du\,dv}{uv}e^{-\frac12(u^2+v^2)}\sinh(\rho uv).
\end{equation}
We now use the known integral
\begin{equation}
\int_0^{\infty}\int_0^{\infty}\frac{du\,dv}{uv}e^{-\frac12(u^2+v^2)}\sinh(\rho uv)=\frac{\pi}{2}\arcsin\rho, \label{eq:integral}
\end{equation}
which can be derived by switching to polar coordinates $u=r\cos\theta$, $v=r\sin\theta$, noting that $\frac{du\,dv}{uv}=\frac{2\,dr\,d\theta}{r\sin2\theta}$, and then using the Frullani integral. The detailed steps are as follows:
\begin{align}
\int_0^{\infty}\int_0^{\infty}\frac{du\,dv}{uv}e^{-\frac12(u^2+v^2)}\sinh(\rho uv)
&= 2\int_0^{\pi/2}\frac{d\theta}{\sin2\theta}\int_0^{\infty}\frac{dr}{r}e^{-r^2/2}\sinh\!\Big(\frac{\rho r^2}{2}\sin2\theta\Big).
\end{align}
Let $t=r^2/2$, then $r\,dr=dt$, so $\frac{dr}{r}=\frac{dt}{2t}$, and the inner integral becomes $\frac12\int_0^{\infty}\frac{dt}{t}e^{-t}\sinh(2a t)$ with $a=\rho\sin2\theta/2$. Using $\sinh(2at)=\frac{e^{2at}-e^{-2at}}{2}$, we get $\frac14\int_0^{\infty}\frac{dt}{t}(e^{-(1-2a)t}-e^{-(1+2a)t})=\frac14\ln\frac{1+2a}{1-2a}$. Hence the double integral reduces to $\frac12\int_0^{\pi/2}\frac{d\theta}{\sin2\theta}\ln\frac{1+\rho\sin2\theta}{1-\rho\sin2\theta}$. The $\theta$ integral yields $\pi\arcsin\rho$, giving the result $\frac{\pi}{2}\arcsin\rho$. Therefore
\begin{equation}
\braket{0|\operatorname{sign}(X)\operatorname{sign}(Y)|0} = \frac{2}{\pi}\arcsin\rho. \label{eq:arcsineres}
\end{equation}
Replacing $\rho = H(f,g)/(\sigma_X\sigma_Y)$ gives the arcsine law
\begin{equation}
\braket{0|\operatorname{sign}(\phi(f))\operatorname{sign}(\phi(g))|0} = \frac{2}{\pi}\arcsin\!\left(\frac{H(f,g)}{\sqrt{H(f,f)H(g,g)}}\right). \label{eq:arcsine}
\end{equation}
This is the fundamental relation used throughout.

\subsection{Hadamard function for a massive scalar field}

In 1+1 dimensions, the vacuum Wightman function for a massive scalar field is
\begin{equation}
W(t,\Delta x)\equiv\bra{0}\phi(t,\Delta x)\phi(0,0)\ket{0} = \frac{1}{2\pi}K_0\!\bigl(m\sqrt{(\Delta x)^2-t^2}\bigr),\qquad |t|<\Delta x, \label{eq:wightman}
\end{equation}
where $K_0$ is the modified Bessel function of the second kind. In the timelike region $|t|>|\Delta x|$, the Wightman function takes the form
\begin{equation}
W(t,\Delta x)=-\frac{1}{4}Y_0\!\bigl(m\sqrt{t^2-(\Delta x)^2}\bigr)-\frac{i}{4}J_0\!\bigl(m\sqrt{t^2-(\Delta x)^2}\bigr),\qquad |t|>|\Delta x|, \label{eq:wightman_timelike}
\end{equation}
where $Y_0$ and $J_0$ are the Neumann and Bessel functions of the first kind, respectively. The Hadamard function, defined as the real part of the Wightman function, is therefore
\begin{equation}
H_m(t,\Delta x)=\frac{1}{2\pi}K_0\!\bigl(m\sqrt{(\Delta x)^2-t^2}\bigr),\qquad |t|<\Delta x, \label{eq:Hm}
\end{equation}
in the spacelike region, and
\begin{equation}
H_m(t,\Delta x)=-\frac{1}{4}Y_0\!\bigl(m\sqrt{t^2-(\Delta x)^2}\bigr),\qquad |t|>|\Delta x|, \label{eq:Hm_timelike}
\end{equation}
in the timelike region. For our spacelike configuration, we always have $|t|<\Delta x$ because $L>\tau+2\tau_0$ ensures $L>|w-\tau|$ for $w\in[-2\tau_0,2\tau_0]$. Hence Eq.~\eqref{eq:Hm} is valid in all integrals except when explicitly evaluating the self-correlation at $L=0$.

\subsection{Explicit expression for $H(f_0,f_\tau)$}

We start from the definition
\begin{equation}
H(f_0,f_\tau)=\frac12\braket{0|\{\phi(f_0),\phi(f_\tau)\}|0}
= \iint dt\,dt'\,\chi(t)\,\chi(t'-\tau)\,H_m(t-t',L). \label{eq:Hdouble}
\end{equation}
Introduce the variables $s=t$, $u=t'-\tau$. Then $t'=u+\tau$, $dt'=du$, and the integral becomes
\begin{equation}
H(f_0,f_\tau)=\iint ds\,du\,\chi(s)\,\chi(u)\,H_m(s-u-\tau,L). \label{eq:Hconv0}
\end{equation}
Define the autocorrelation function
\begin{equation}
K(w)=\int_{-\infty}^{\infty}\chi(s)\chi(s-w)\,ds. \label{eq:Kdef}
\end{equation}
By exchanging the order of integration and setting $w=s-u$, we obtain the convolution form
\begin{equation}
H(f_0,f_\tau)=\int_{-\infty}^{\infty} K(w)\,H_m(w-\tau,L)\,dw. \label{eq:Hconv}
\end{equation}
For the rectangular window $\chi(t)=1$ for $|t|<\tau_0$ (we set $\tau_0=1$ henceforth), we have
\begin{equation}
K(w)=\begin{cases}
2-|w|, & |w|<2,\\
0, & |w|\ge2.
\end{cases} \label{eq:Krect}
\end{equation}
Inserting Eq.~\eqref{eq:Hm} into Eq.~\eqref{eq:Hconv} and using the evenness of $K$ and $H_m$ (since $K(w)=K(-w)$ and $H_m(w-\tau,L)=H_m(\tau-w,L)$), we obtain
\begin{align}
H(\tau,L)&\equiv H(f_0,f_\tau)\nonumber\\
&=\int_{-2}^{2}(2-|w|)\,H_m(w-\tau,L)\,dw\nonumber\\
&=\frac{1}{2\pi}\int_0^{2}(2-w)\Bigl[K_0\!\bigl(m\sqrt{L^2-(\tau+w)^2}\bigr)+K_0\!\bigl(m\sqrt{L^2-(\tau-w)^2}\bigr)\Bigr]dw. \label{eq:Htau}
\end{align}
Similarly, the self-correlation $H_0\equiv H(f_0,f_0)$ is obtained by setting $\tau=0$ and $L=0$ in Eq.~\eqref{eq:Hconv}:
\begin{equation}
H_0 = \int_{-\infty}^{\infty} K(w)\,H_m(w,0)\,dw. \label{eq:H0step1}
\end{equation}
For equal-space points ($L=0$), the two spacetime arguments are timelike separated (except at $w=0$), so the kernel $H_m(w,0)$ cannot be taken from the spacelike formula Eq.~\eqref{eq:Hm}. Instead, we use the timelike expression Eq.~\eqref{eq:Hm_timelike} with $\Delta x=0$:
\begin{equation}
H_m(w,0)=-\frac{1}{4}Y_0(m|w|). \label{eq:Hm_timelike_zero}
\end{equation}
Substituting Eq.~\eqref{eq:Hm_timelike_zero} and the explicit form of $K(w)$ from Eq.~\eqref{eq:Krect} into Eq.~\eqref{eq:H0step1},
and using the evenness of $K(w)$ and $Y_0$, we obtain
\begin{equation}
H_0 = 2\int_0^{2}(2-w)\,H_m(w,0)\,dw = -\frac{1}{2}\int_0^{2}(2-w)\,Y_0(mw)\,dw. \label{eq:H0}
\end{equation}
The integral is well-defined and finite because $Y_0(mw)\sim (2/\pi)\ln(mw/2)$ as $w\to0^+$, which is integrable against $(2-w)$.

\subsection{Correlation function and LGI parameter}

From the arcsine law Eq.~\eqref{eq:arcsine} and the fact that $H(f_0,f_0)=H(f_\tau,f_\tau)=H_0$, we get
\begin{equation}
C(\tau,L)=\frac{2}{\pi}\arcsin\!\left(\frac{H(\tau,L)}{H_0}\right). \label{eq:C}
\end{equation}
Then the LGI parameter is given by Eq.~\eqref{eq:K3def2} with $C(2\tau,2L)$ obtained from the same formula by replacing $\tau\to2\tau$, $L\to2L$.

\section{Numerical results}
\label{sec:numerics}

We set $\tau_0=1$ and use the rectangular window. For each $\tau$ in the overlapping region $(0,2)$, we take $L=\tau+2+\epsilon$ with $\epsilon=10^{-3}$ to satisfy the spacelike condition.
We compute $H(\tau,L)$ and $H_0$ numerically using adaptive integration (Mathematica's \texttt{NIntegrate}) and then evaluate $K_3(\tau,L)$ from Eqs.~\eqref{eq:C} and \eqref{eq:K3def2}.

The calculation is performed for two sets of masses: $m=0.1$, $0.2$, $0.3$, $0.4$ (heavier masses) and
$m=0.001$, $0.002$, $0.003$, $0.004$ (light masses), to fully probe the mass dependence down to the nearly massless limit.

\begin{figure}[htb]
\centering
\includegraphics[width=0.8\linewidth]{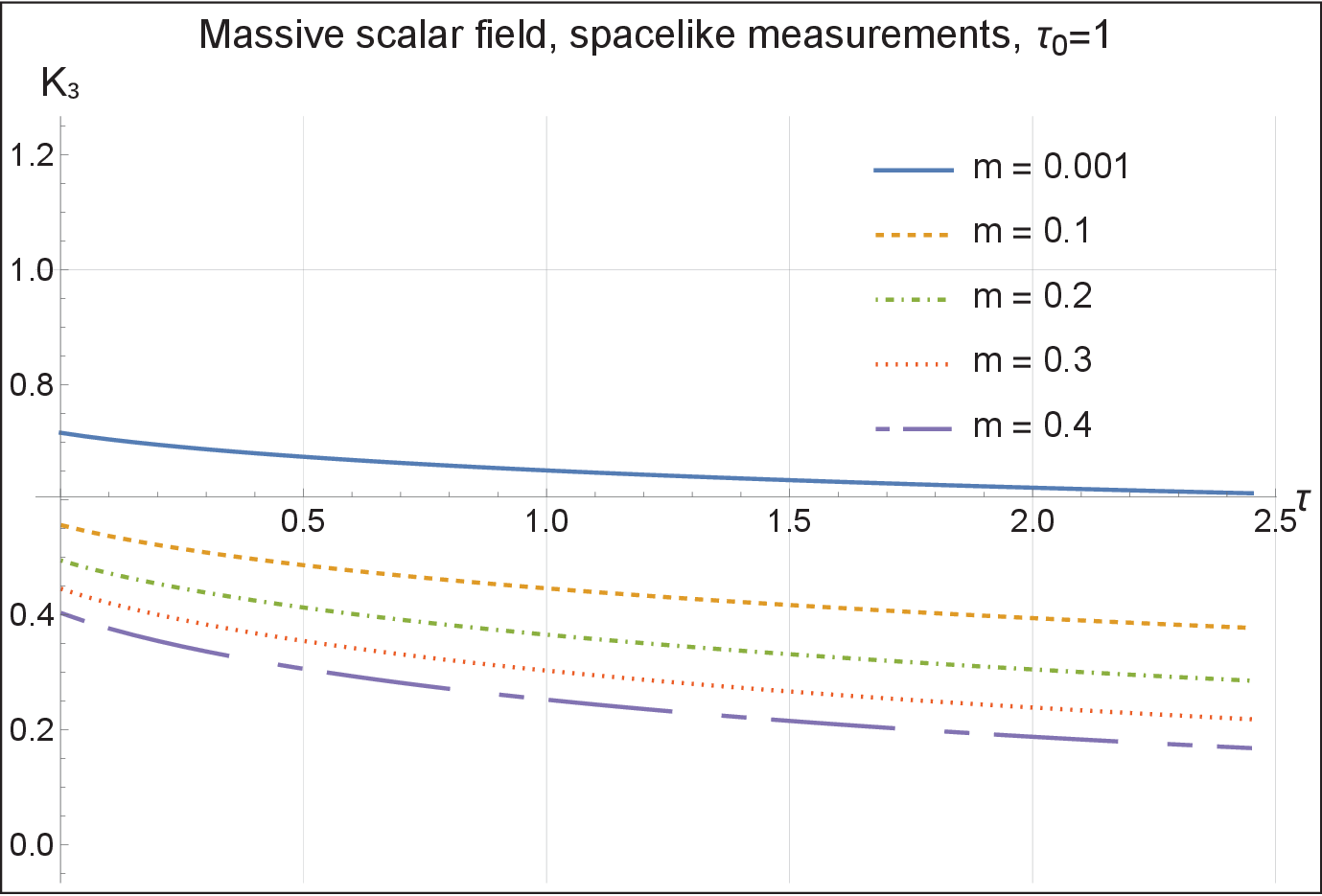}
\caption{$K_3(\tau,L)$ for the massive scalar field with spacelike-separated measurements ($L=\tau+2+\epsilon$, $\epsilon=10^{-3}$) for various masses $m$.
The thin horizontal line marks the classical bound $K_3=1$; for all masses considered, $K_3$ remains strictly below $1$.
For each mass, the maximum occurs at the smallest accessible $\tau$ ($\tau=0.001$), and $K_3^{\max}$ decreases monotonically as $m$ increases, from $K_3^{\max}\approx0.716$ at $m=0.001$ down to $0.403$ at $m=0.4$.
For $\tau\ge2$, the two time windows no longer overlap; the curves decay exponentially with $\tau$, as expected from the spacelike decay of the massive Hadamard function.}
\label{fig:K3}
\end{figure}

For $\tau \ge 2$, the two time windows no longer overlap. In this regime, the Hadamard inner product does not vanish identically; rather, it is controlled by the spacelike decay of $K_0(m\sqrt{L^2-(w-\tau)^2})$. For a massive field, this decay is exponential, as can be seen in Fig.~\ref{fig:K3} for $\tau > 2$. In the massless limit $m\to0$, the decay becomes algebraic, specifically $\sim 1/\tau$ for large $\tau$ in $1+1$ dimensions.

Fig.~\ref{fig:K3} shows $K_3(\tau,L)$ as a function of $\tau$ for these masses. Table~\ref{tab:max} lists the maximum value of $K_3$ for each mass and the corresponding $\tau_{\text{max}}$.

\begin{table}[htb]
\centering
\caption{Maximum $K_3$ and its location $\tau_{\text{max}}$ for different masses.}
\begin{tabular}{c|c|c}
$m$ & $K_3^{\max}$ & $\tau_{\text{max}}$ \\ \hline
0.001 & 0.7162 & 0.001 \\
0.002 & 0.7028 & 0.001 \\
0.003 & 0.6940 & 0.001 \\
0.004 & 0.6873 & 0.001 \\
0.1   & 0.5558 & 0.001 \\
0.2   & 0.4942 & 0.001 \\
0.3   & 0.4450 & 0.001 \\
0.4   & 0.4029 & 0.001 \\
\end{tabular}
\label{tab:max}
\end{table}

The numerical results clearly demonstrate that $K_3$ is always less than $1$ for all masses studied. The maximum value of $K_3$ is approximately $0.716$, obtained at the smallest mass $m=0.001$. As the mass increases, $K_3^{\max}$ decreases monotonically, approaching zero for large $m$. Even in the nearly massless limit $m\to0$, $K_3$ saturates at a value well below the classical bound. The peak always occurs at the smallest accessible $\tau$ ($\tau=0.001$ in our numerical grid), consistent with the physical expectation that the strongest correlations appear when the measurement windows have maximal overlap.

Thus, under strict noninvasive (spacelike-separated) measurements, the vacuum state of a free massive scalar field does \emph{not} violate the Leggett-Garg inequality. This result is robust across the entire mass range, from the nearly massless regime ($m=0.001$) to the heavy-mass regime ($m=0.4$).

\section{Conclusion}
\label{sec:conclusion}

We have studied the Leggett-Garg inequality in the vacuum state of a free massive scalar field using the dichotomic observable $\operatorname{sign}(\phi(f))$.
Our protocol, in which three observers at rest at distinct spatial points perform local measurements on three independent ensembles of the vacuum state, ensures that the noninvasive measurability condition --- the central experimental obstacle in all LGI tests --- is rigorously satisfied through spacelike separation.
We showed that for non-overlapping time windows the correlation function is exponentially suppressed for a massive field, consistent with the cluster property of the vacuum, rather than vanishing identically.
For overlapping windows, our numerical calculations for a rectangular time window ($\tau_0=1$) across masses from $m=0.001$ to $m=0.4$ establish that $K_3$ remains strictly below the classical bound for all masses, with a monotonic suppression as the mass increases.

This negative result acquires its full physical significance when contrasted with the well-known maximal violation of the Bell-CHSH inequality by the same vacuum state.
The asymmetry has a sharp origin: spacelike Bell violations probe the non-factorizable entanglement structure of the vacuum across causally disconnected regions --- a genuinely quantum feature without any classical analogue.
Temporal LGI tests, however, probe correlations along a single worldline, where the dynamics are governed by the Klein-Gordon equation.
Crucially, the vacuum is a Gaussian state, and the free-field Heisenberg evolution preserves this Gaussianity.
Gaussian states are known to saturate certain uncertainty relations and exhibit limited temporal quantumness, which naturally precludes violations of the macrorealism bound.
Our negative result thus faithfully reflects not a failure of the vacuum's nonclassicality, but rather a fundamental boundary of its manifestation: macrorealism remains intact along the worldlines of free Gaussian fields.

This distinction raises a compelling question for future investigations: do interactions or non-Gaussianity enable LGI violations?
In interacting QFTs, the nonlinear mode couplings can generate non-Gaussian correlations that may break the classical bound; likewise, engineered non-Gaussian states --- such as cat states or squeezed number states --- possess higher-order correlations that could potentially yield temporal violations.
Our spacelike-separated protocol, by rigorously satisfying NIM, provides the ideal framework to address these questions.
More broadly, the protocol can be applied to other observables (e.g., the parity operator or smeared projectors) and to other states, offering a rigorous methodological benchmark for future studies of macrorealism in relativistic quantum systems.
Extensions to higher dimensions, interacting theories, and non-Gaussian states, as well as a deeper investigation of the mass-induced suppression of $K_3$ in relation to the field's correlation length, constitute natural next steps.

\acknowledgments

This work is supported by the National Natural Science Foundation of China under Grant No. 11005031.

\bibliographystyle{apsrev4-1}
\bibliography{lgimsv}

\end{document}